\documentclass[10pt,letterpaper]{article}
\usepackage[top=0.85in,left=2.75in,footskip=0.75in]{geometry}

\usepackage{amsmath,amssymb}

\usepackage{changepage}

\usepackage{tabularx}

\usepackage[utf8x]{inputenc}

\usepackage{textcomp,marvosym}

\usepackage{cite}

\usepackage{nameref,hyperref}

\hypersetup{
    colorlinks=true,
    linkcolor=blue,
    filecolor=magenta,      
    urlcolor=cyan,
}

\usepackage{microtype}
\DisableLigatures[f]{encoding = *, family = * }

\usepackage[table]{xcolor}

\usepackage{array}

\newcolumntype{+}{!{\vrule width 2pt}}

\newlength\savedwidth

\raggedright
\usepackage[aboveskip=1pt,labelfont=bf,labelsep=period,justification=raggedright,singlelinecheck=off]{caption}

\makeatletter
\renewcommand{\@biblabel}[1]{\quad#1.}
\makeatother

\usepackage{lastpage,fancyhdr,graphicx}
\usepackage{epstopdf}
\fancyheadoffset[L]{2.25in}
\fancyfootoffset[L]{2.25in}
\begin{document}

\pagebreak
\hspace{0pt}
\vfill
\vspace*{0.2in}

\begin{flushleft}
{\Large
\textbf\newline{Spikebench: an open benchmark for spike train time-series classification} 
}
\newline
\\
Ivan Lazarevich\textsuperscript{1*},
Ilya Prokin\textsuperscript{2},
Boris Gutkin\textsuperscript{1,3\Yinyang},
Victor Kazantsev\textsuperscript{4,5\Yinyang},
\\
\bigskip
\textbf{1} \'Ecole Normale Sup\'erieure, Laboratoire de Neurosciences Cognitives, Group for Neural Theory, Paris, France
\\
\textbf{2} Coeus Metis Labs, Bordeaux, France
\\
\textbf{3} Center for Cognition and Decision Making, National Research University Higher School of Economics, Moscow, Russia
\\
\textbf{4} Neurotechnology Department, Lobachevsky State University of Nizhny Novgorod, Nizhny Novgorod, Russia
\\
\textbf{5} Baltic Center for Artificial Intelligence and Neurotechnology, Immanuel Kant Baltic Federal University, Kaliningrad, Russia

\bigskip

%
%
\Yinyang These authors contributed equally to this work.





* ivan@lazarevi.ch

\end{flushleft}
\vfill
\hspace{0pt}
\pagebreak

\clearpage
\section*{Abstract}
Modern well-performing approaches to neural decoding are based on machine learning models such as decision tree ensembles and deep neural networks. The wide range of algorithms that can be utilized to learn from neural spike trains, which are essentially time-series data, results in the need for diverse and challenging benchmarks for neural decoding, similar to the ones in the fields of computer vision and natural language processing. In this work, we propose a spike train classification benchmark, based on open-access neural activity datasets and consisting of several learning tasks such as stimulus type classification, animal's behavioral state prediction, and neuron type identification. We demonstrate that an approach based on hand-crafted time-series feature engineering establishes a strong baseline performing on par with state-of-the-art deep learning-based models for neural decoding. We release the \href{https://github.com/lzrvch/spikebench}{code allowing to reproduce the reported results}.

\section*{Author summary}
Machine learning-based neural decoding has been shown to outperform traditional approaches like Wiener and Kalman filters on certain key tasks. To further the advancement of neural decoding models, such as improvements in deep neural network architectures and better feature engineering for classical ML models, there need to exist common evaluation benchmarks similar to the ones in the fields of computer vision or natural language processing. In this work, we propose a benchmark consisting of several \textit{individual neuron} spike train classification tasks based on open-access data from a range of animals and brain regions. We demonstrate that it is possible to achieve meaningful results in such a challenging benchmark using the massive time-series feature extraction approach, which is found to perform similarly to state-of-the-art deep learning approaches.


\section*{Introduction}

The latest advances in multi-neuronal recording technologies such as two-photon calcium imaging \cite{pachitariu2016suite2p}, extracellular recordings with multi-electrode arrays \cite{tsai2015high}, Neuropixels probes \cite{steinmetz2018challenges} allow producing large-scale single-neuron resolution brain activity data with remarkable magnitude and precision. Some of the neural spiking data recorded in animals have been released to the public in the scope of data repositories such as CRCNS.org \cite{teeters2009crcns}. In addition to increasing experimental data access, various neural data analysis tools have been developed, in particular for the task of neural decoding, which is often posed as a supervised learning problem \cite{glaser2017machine}: given the firing activity of a population of neurons at each time point, one has to predict the value of a certain quantity pertaining to animal's behavior such as its velocity at a given point in time. 

Such a formulation of the neural decoding task implies that it is a multivariate time-series regression or classification problem. An array of supervised learning methods focused specifically on general time-series data has been developed over the years, ranging from classical approaches \cite{bagnall2017great} to deep neural networks for sequential data \cite{fawaz2019deep}. It is not fully clear, however, how useful these methods are for the specific tasks of learning from neural spiking data. In order to establish a sensible ranking of these algorithms for neural decoding, there is a need for a common spiking activity recognition benchmark. In this work, we propose a diverse and challenging spike train classification benchmark based on several open-access neuronal activity datasets. This benchmark incorporates firing activity from different brain regions of different animals (retina, prefrontal cortex, motor, and visual cortices) and comprises distinct task types such as visual stimulus type classification, animal's behavioral state prediction from individual spike trains, and interneuron subtype recognition from firing patterns. All of these tasks are formulated as univariate time-series classification problems, that is, one needs to predict the target category based on an individual spike train chunk recorded from a single neuron. The formulation of the classification problems implies that the predicted category is stationary across the duration of the given spike train sample.

Our main contributions can be summarized as follows:

\begin{itemize}
    \item We propose a diverse spike train classification benchmark based on open-access data.
    \item We show that global information such as the animal's behavioral state or stimulus type can be decoded (with high accuracy) from \textit{single-neuron} spike trains containing several tens of interspike intervals.
    \item We establish a strong baseline for spike train classification based on hand-crafted time-series feature engineering that performs on par with state-of-the-art deep learning models.
\end{itemize}

Well-established machine learning techniques such as gradient-boosted decision tree ensembles and recurrent neural networks have been successfully applied both to neural activity decoding (predicting stimuli/action from spiking activity) \cite{glaser2017machine, livezey2021deep} as well as neural encoding (predicting neural activity from stimuli) \cite{benjamin2018modern}. Neural decoding tasks are often formulated as regression problems, wherein binned spiking count time series of a single fixed neural population are used to predict the animal's position or velocity in time.

A number of previous studies on feature vector representations of spike trains also focused on defining a spike train distance metric \cite{tezuka2018multineuron} for identification of neuronal assemblies \cite{humphries2011spike}. Several different definitions of the spike train distance exist such as van Rossum distance \cite{rossum2001novel}, Victor-Purpura distance \cite{victor1997metric}, SPIKE- and ISI- synchronization distances \cite{mulansky2016pyspike} (for a thorough list of existing spike train distance metrics see \cite{tezuka2018multineuron}). These distance metrics were used to perform spike train clustering and classification based on the k-Nearest-Neighbors approach \cite{tezuka2015spike}. Jouty et al. \cite{jouty2018non} employed ISI and SPIKE distance measures to perform clustering of retinal ganglion cells based on their firing responses to a given stimulus.

In addition to characterization with spike train distance metrics, some previous works relied on certain statistics of spike trains to differentiate between cell types. Charlesworth et al. \cite{charlesworth2015quantitative} calculated basic statistics of multi-neuronal activity from cortical and hippocampal cultures and were able to perform clustering and classification of activity between these culture types. Li et al. \cite{li2015computational} used two general features of the interspike interval (ISI) distribution to perform clustering analysis to identify neuron subtypes. Such approaches represent neural activity (single or multi-neuron spiking patterns) in a low-dimensional feature space where the hand-crafted features are defined to address specific problems and might not provide an optimal feature representation of spiking activity data for a general decoding problem.
Finally, not only spike timing information can be used to characterize neurons in a supervised classification task. Jia et al. \cite{jia2018high} used waveform features of extracellularly recorded action potentials to classify them by brain region of origin.

The aforementioned works were aimed at, to some extent or another, trying to decode the properties of neurons or stimuli given recorded spiking data. In some of the cases, the datasets used were not released to be openly available, and in some of the cases, the predictive models used constituted quite simple baselines for the underlying decoding/cell identification tasks. In this work, we aim to propose a benchmark based on open-access datasets that is diverse and challenging enough to robustly demonstrate gains of advanced time-series machine learning approaches as compared to some of the simple baselines used in previous works. We release the \href{https://github.com/lzrcvh/spikebench}{code allowing to reproduce the reported results}.

\section*{Materials and methods}
\subsection*{Overview of time series classification methods}

We applied general time series feature representation methods \cite{bagnall2017great} for the classification of neuronal spike train data. Most approaches in time series classification are focused on transforming the raw time series data into an effective feature space representation before training and applying a machine learning classification model. Here we give a brief overview of state-of-the-art approaches one could utilize in order to transform time series data into a feature vector representation for efficient neural activity classification.
\begin{figure*}
    \centering
    \includegraphics[width=\textwidth]{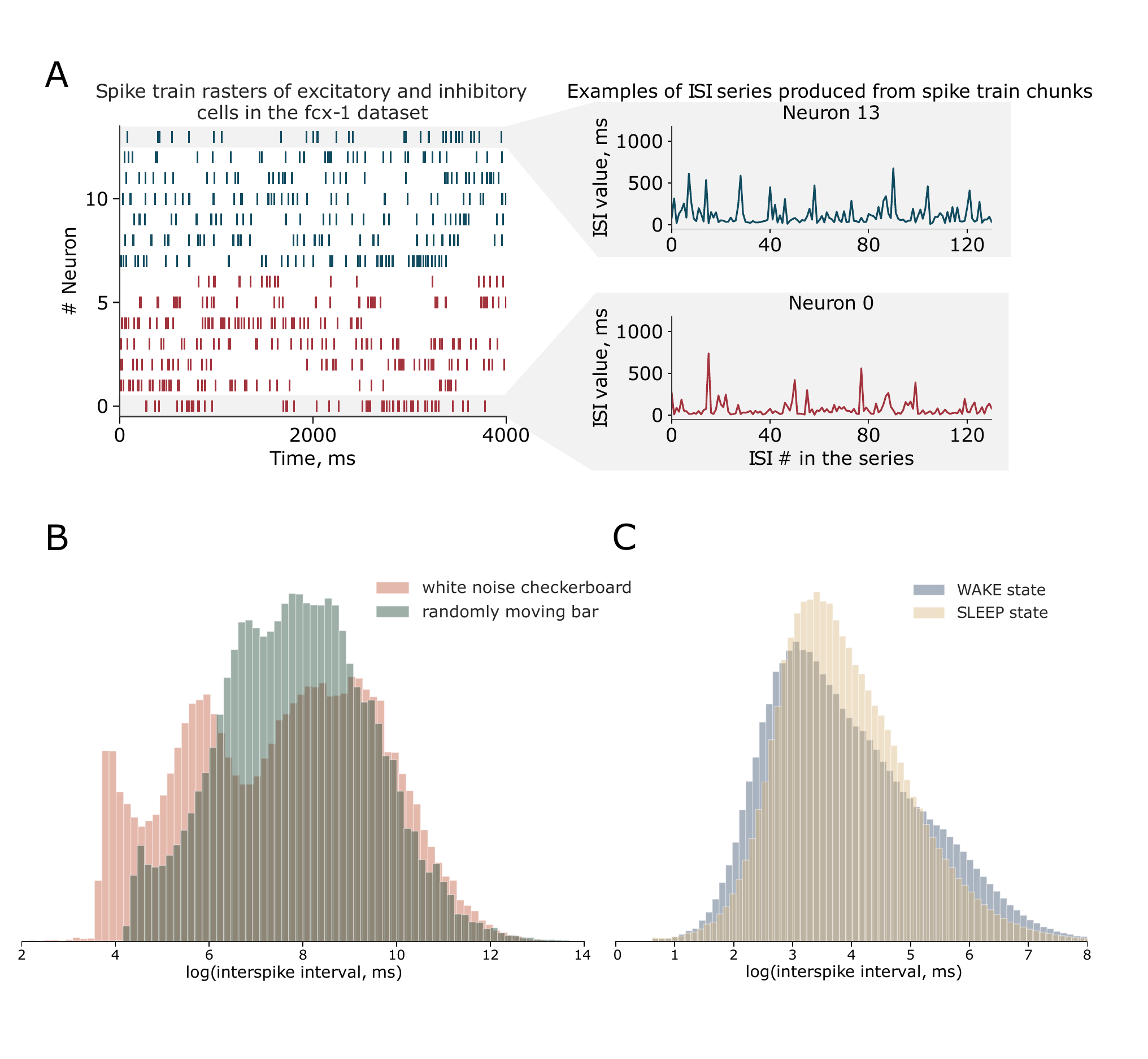}
    \caption{(A) Examples of spiking activity recordings in the CRCNS fcx-1 dataset in the WAKE state. Left: spike train raster of a random subset of excitatory cells (red) and inhibitory cells (blue). Right: examples of ISI series produced from spike train chunks of inhibitory/excitatory cells in the fcx-1 dataset. (B) Interspike interval value distribution histograms generated from the aggregated spike trains of retinal ganglion cells in response to a ``white noise checkerboard'' visual stimulus (red) and a ``randomly moving bar'' stimulus (green). (C) Interspike interval value distribution histograms generated from the aggregated PFC spike trains (fcx-1 dataset) corresponding to the WAKE (blue) or SLEEP (yellow) state of the rat.}
    \label{fig:figure1}
\end{figure*}

\subsubsection*{Neighbor-based models with time series distance measures}

A strong baseline algorithm for time series classification is k-nearest-neighbors (kNN) with a suitable time series distance metric such as the Dynamic Time Warping (DTW) distance or the edit distance (ED) \cite{bagnall2017great}. In this work, we evaluated the performance of nearest-neighbor models for generic distance measures such as $l_p$ and DTW distance, converting spike trains to the interspike-interval (ISI) time-series representation prior to calculating the spike-train distances. 
Some of the distance metrics we also used for evaluation are essentially distribution similarity measures (e.g. Kolmogorov-Smirnov distance, Earth mover's distance) which allow comparing ISI value distributions within spike trains. Such a spike train distance definition would only use the information about the ISI distribution in the spike train, but not about its temporal structure.
Alternatively, one can keep the original event-based representation of the spike train and compute the spike train similarity metrics such as van Rossum or Victor-Purpura distances or ISI/SPIKE distances \cite{tezuka2018multineuron}.

The choice of the distance metric determines which features of the time series are considered as important. Instead of defining a complex distance metric, one can explicitly transform time series into a feature space by calculating various properties of the series that might be important (e.g. mean, variance). After assigning appropriate weights to each feature one can use kNN with any standard distance metric. Moreover, such a representation allows the application of any state-of-the-art machine learning classification algorithm beyond kNN to obtain better classification results. In the following, we discuss approaches using various feature space representations available for time series data.

\subsubsection*{Models using hand-crafted time series features}

One of the useful and intuitive approaches in time series classification is focused on manually calculating a set of descriptive features for each time series (e.g. their basic statistics, spectral properties, other measures used in signal processing, and so on) and using these feature sets as vectors describing each sample series. There exist approaches that enable automated calculation of a large number of time series features which may be typically considered in different application domains. Such approaches include automated time series phenotyping implemented in the \textit{hctsa} MATLAB package \cite{fulcher2017hctsa} and automated feature extraction in the \textit{tsfresh} Python package \cite{christ2018time}. Here we utilize the \textit{tsfresh} package which enables the calculation of 779 descriptive time series features for each spike train, ranging from Fourier and wavelet expansion coefficients to coefficients of a fitted autoregressive process.

Once each time series (spike train) is represented as a feature vector, the spiking activity dataset has the standard form of a matrix with size $[n_{\text{samples}}, n_{\text{features}}]$ rather than the raw dataset with shape $[n_{\text{samples}}, n_{\text{timestamps}}]$. This standardized dataset can be then used as an input to any machine learning algorithm such as logistic regression or gradient boosted trees \cite{friedman2001greedy}. We found this approach to set a strong baseline for all of the classification tasks we considered.

\subsubsection*{Deep learning models}

Lastly, there are deep learning-based approaches working well for time-series classification \cite{fawaz2019deep} such as deep recurrent networks like LSTMs and GRUs \cite{karim2019multivariate} and 1D convolutional neural networks (1D-CNNs) \cite{zhao2017convolutional,fawaz2020inceptiontime}. While rather generic model architectures have been typically applied to neural decoding tasks \cite{glaser2017machine, livezey2021deep}, there exist models specifically designed for time-series classification and regression tasks, like InceptionTime \cite{fawaz2020inceptiontime}, achieving state-of-the-art results on benchmarks like the UCR Time Series Classification Archive \cite{dau2019ucr}. Recent developments in deep learning models for time series also include the Time Series Transformer \cite{zerveas2020transformer} and convolutional architectures like the Omniscale-CNN \cite{tang2020rethinking}. Perhaps surprisingly, we found that deep learning models could not significantly outperform the baseline with hand-crafted time series features on the spike train classification tasks, oftentimes performing worse than the baseline. 
\subsection*{The proposed spike train classification benchmark.}

We propose a spike train classification benchmark comprising several different open-access datasets and distinct classification tasks. The datasets used for the benchmark are as follows:

\begin{itemize}
\item \textbf{Retinal ganglion cell stimulus type classification} based on the published dataset \cite{prentice2016error,RetinalDataset}: Spike time data from multi-electrode array recordings of salamander retinal ganglion cells under four stimulus conditions: a white noise checkerboard, a repeated natural movie, a non-repeated natural movie, and a bar exhibiting random one-dimensional motion. We define the 4-class classification task to predict the stimulus type given the spike train chunk, also considering binary classification tasks for pairs of stimuli types (e.g. ``white noise checkerboard'' vs. ``randomly moving bar'').

\item \textbf{WAKE/SLEEP classification} based on fcx-1 dataset \cite{fcx1dataset, fcx-buszaki} from \href{http://crcns.org/}{CRCNS.org}\cite{teeters2009crcns}: Spiking activity and Local-Field Potential (LFP) signals recorded extracellularly from frontal cortices of male Long Evans rats during wake and sleep states without any particular behavior, task or stimulus. Around 1100 units (neurons) were recorded, ~120 of which are putative inhibitory cells and the rest is putative excitatory cells. Fig \ref{fig:figure1} shows several examples of spiking activity recordings that can be extracted from the fcx-1 dataset. The authors classified cells into an inhibitory or excitatory class based on the action potential waveform (action potential width and peak time). Sleep states (SLEEP activity class) were labeled semi-automatically based on extracted LFP and electromyogram features, and the non-sleep state was labeled as the WAKE activity class. We define the binary classification task as the prediction of WAKE or SLEEP animal state given a spike train chunk recorded from a putative excitatory cell.

\item \textbf{Interneuron subtype classification task} based on the Allen Cell Types dataset \cite{AllenCellTypesDataset}: Whole cell patch clamp recordings of membrane potential in neurons of different types. We selected the PV, VIP, and SST interneurons from the whole dataset, as these interneuron groups comprise the majority of inhibitory cells in the prefrontal cortex \cite{rudy2011three}. We selected the spike trains recorded under the naturalistic noise stimulation protocol (as a proxy for the \textit{in vivo} spontaneous activity in these cells). The non-trivial prediction task is defined for VIP vs. SST spike train classification since the PV interneuron spike trains can be easily distinguished from the other interneuron types. The latter is due to a significantly higher firing frequency in PV interneurons that we found in the Allen Cell Types dataset.

\item \textbf{Unsupervised temporal structure recognition task.} We defined a set of spike train classification tasks constructed in a self-supervised manner \cite{jing2020self}. In such tasks, we take any set of (unlabelled) neuronal spike train recordings and generate an additional set of spike trains by applying a given transformation to the original data. The target classification task is to determine whether a given spike train chunk belongs to the original dataset or to the transformed one. Note that this task can be constructed for any spiking dataset without the need for ground truth labels, i.e. in an unsupervised way. The spike train transformations we consider here are (i) adding spike timing jitter via, in particular, timing noise following a truncated normal distribution, (ii) random shuffling of the interspike intervals in the spike train, (iii) reversing the spike train. The models trained in such tasks learn to detect the temporal structure of the original spike trains since the order/precise values of interspike intervals have been disrupted by the transformation (e.g. by ISI shuffling), while the ISI value distribution is preserved by some of the transformations (e.g. by the shuffling and reversal operations). The final trained model accuracy in a shuffled vs. non-shuffled spike train classification task can thus be thought of as a measure of temporal structure in the original spiking dataset (test set accuracy would be on the chance level if the ISI values in the original spike trains were independently sampled from a fixed value distribution, i.e. the exact ordering of the ISIs did not contain any predictive information). We consider the fcx-1 and retinal ganglion cell datasets described above to construct the temporal structure recognition tasks.

\end{itemize}

\subsection*{Validation scheme and data preprocessing}

Suppose we are given a dataset containing data from several animals each recorded multiple times with a large number of neurons captured in each recording. For each recorded neuron, we have a corresponding spike train captured over a certain period of time (assuming that the preprocessing steps like spike sorting or spiking time inference from fluorescence traces were performed beforehand). The number of spikes within each spike train is going to be variable. A natural way to standardize the length of spike-train sequences would be dividing the full spike train into chunks of $N$ spike times, where $N$ is fixed for each chunk. We produce these spike train chunks for all datasets by moving a sliding window of a fixed number of spikes across each single-neuron spike train. Each window will thus contain the same number of spikes (and hence ISIs) but vary in time duration. To summarize, the preprocessing pipeline for all datasets is as follows

\begin{itemize}
    \item Encode all of the spike trains in the interspike interval format (i.e. time-series [$ISI_1$, $ISI_2$, $ISI_3$, ...])
    \item Apply a rolling window of size $N$ with step $S$ to each single-neuron spike train separately, producing several spike train chunks (of $N$ interspike intervals) for each neuron.
    \item Pool all the chunks from all neurons together to form a data matrix of size $[M, N]$ where $M$ is the total number of samples (chunks) in the training/test dataset and $N$ is the chunk size (number of ISIs in the chunk).
    \item Apply a logarithmic transform ($f(x) = log(x + 1)$) to each sample (due to the originally heavy-tail distribution of ISI values in the data) and standard scaling (z-score normalization) to the dataset. Alternatively, encode the spike train chunks using a binned spike count representation.
\end{itemize}

We train a decoder (classifier) using the training set data matrix, so a single decoder is trained for all the neurons in the training set from single-neuron spike train chunks. Regarding the last step of the preprocessing pipeline, we demonstrate that it is crucial when dealing with ISI-encoded data to perform this step to avoid significant accuracy degradation (see Fig \ref{fig:figureS2}). We have applied this preprocessing step to ISI-encoded data in all experiments throughout the study (unless stated otherwise).

The validation strategy we use in this work is based on group splits, which means we determine the split into the training and the test datasets based on animal identifiers available in the original data. The motivation is that in cases recordings are performed in several animals and corresponding animal identifiers are available, the set of animals used to construct the training dataset and the set of animals for the test dataset should not overlap in order to test whether the trained decoding models could generalize across different animals. In case animal identifiers are not available, we split the dataset into training and testing based on non-overlapping neuron identifiers in train and test.

Most of the datasets in the benchmark have an imbalanced class distribution. For performance evaluation, we consider two different setups in this work: (i) keeping all the available data in the training/test datasets, and measuring performance with metrics that are robust to class imbalance, such as the Cohen's kappa score \cite{cohenkappa}, and the geometric mean score \cite{geomean} and (ii) balancing the class distribution in the training and testing datasets by data undersampling and measuring standard classification metrics such as accuracy and AUC-ROC. The second approach allows to rank models without the effect of class imbalance on training and evaluation \cite{imblearn}, however some data points are being lost in the undersampling process. Although AUC-ROC is generally considered to be a performance metric less affected by the class imbalance than e.g. accuracy, we do not report AUC-ROC values for the full (imbalanced) data setting because of its possible skewness \cite{aucimbalance}.

\begin{table*}[!ht]
\centering
\begin{adjustwidth}{-2.25in}{0in}
{\color{black}\begin{tabularx}{\linewidth}{|l|X|X|X|}
\hline
{} & {Retina dataset (Cohen's kappa score)} & {Retina dataset (geometric mean score)} \\
\hline 
{kNN, $k=1$, $l_1$} & {0.3180} & {0.5337}  \\ \hline 
{kNN, $k=5$, $l_1$} & {0.2819} & {0.4705}   \\ \hline
{kNN, $k=5$, $l_2$} & {0.0434} & {0.2328} \\ \hline 
{kNN, $k=1$, DTW ($r=50$)} & {0.4759} & {0.7608}  \\ \hline 
{kNN, $k=1$, Victor-Purpura} & {-0.0184} & {0.4511}  \\ \hline 
{kNN, $k=1$, ISI} & {0.2920} & {0.6611}  \\ \hline 
{kNN, $k=1$, KS} & {{ \bf0.5464}} & {{ \bf 0.8316}}  \\ \hline 
\end{tabularx}}
\vspace{1mm}
\caption{{\color{black}Spike train classification results for the retinal neuron activity dataset for nearest-neighbor models with different distance metrics. The task is defined as binary classification of the stimulus type (``white noise checkerboard'' or ``randomly moving bar'').}}
\label{tab:table1}
\end{adjustwidth}
\end{table*}

\section*{Results}

\subsection*{Visual stimulus type classification from retinal spike trains}

We first start by looking at the retinal ganglion cell spike train classification task. Recorded spike trains in the dataset are associated with one of the four categories corresponding to different visual stimulus types, labeled with ``white noise checkerboard'', ``randomly moving bar'', ``repeated natural movie'' and ``unique natural movie''. The classification task is, given a chunk of the spike train recording, to predict the corresponding stimulus type category. The number of neurons in the dataset belonging to each category is 155, 140, 178, and 152, respectively. The number of interspike intervals is quite variable among individual cells (due to firing rate variability) ranging from 100 ISIs per recording to as much as 60000 ISIs per recording.

We focus on the binary classification task aimed at predicting one of the two types of stimuli: ``randomly moving bar'' (corresponding to a label of 0) or ``white noise checkerboard'' (corresponding to a label of 1). We select recorded spike trains corresponding to those stimuli types and split 70\% of recorded neurons (108 neurons) for the training part of the dataset and the remaining 30\% (46 neurons) for the testing dataset. We encode spike trains using the ISI representation and apply a rolling window of size equal to 200 ISIs with a stride (step) of 100 ISIs to each recorded neuron. This results in 8006 training samples and 3650 testing samples, each sample containing 200 ISIs. The average target value in the training set is 0.7660 and 0.7882 in the testing set, hence class imbalance is present in the retinal dataset. 

\subsubsection*{Nearest-neighbor models for spike train classification}
\label{sec:knn}

\begin{figure*}
    \centering
    \includegraphics[width=\textwidth]{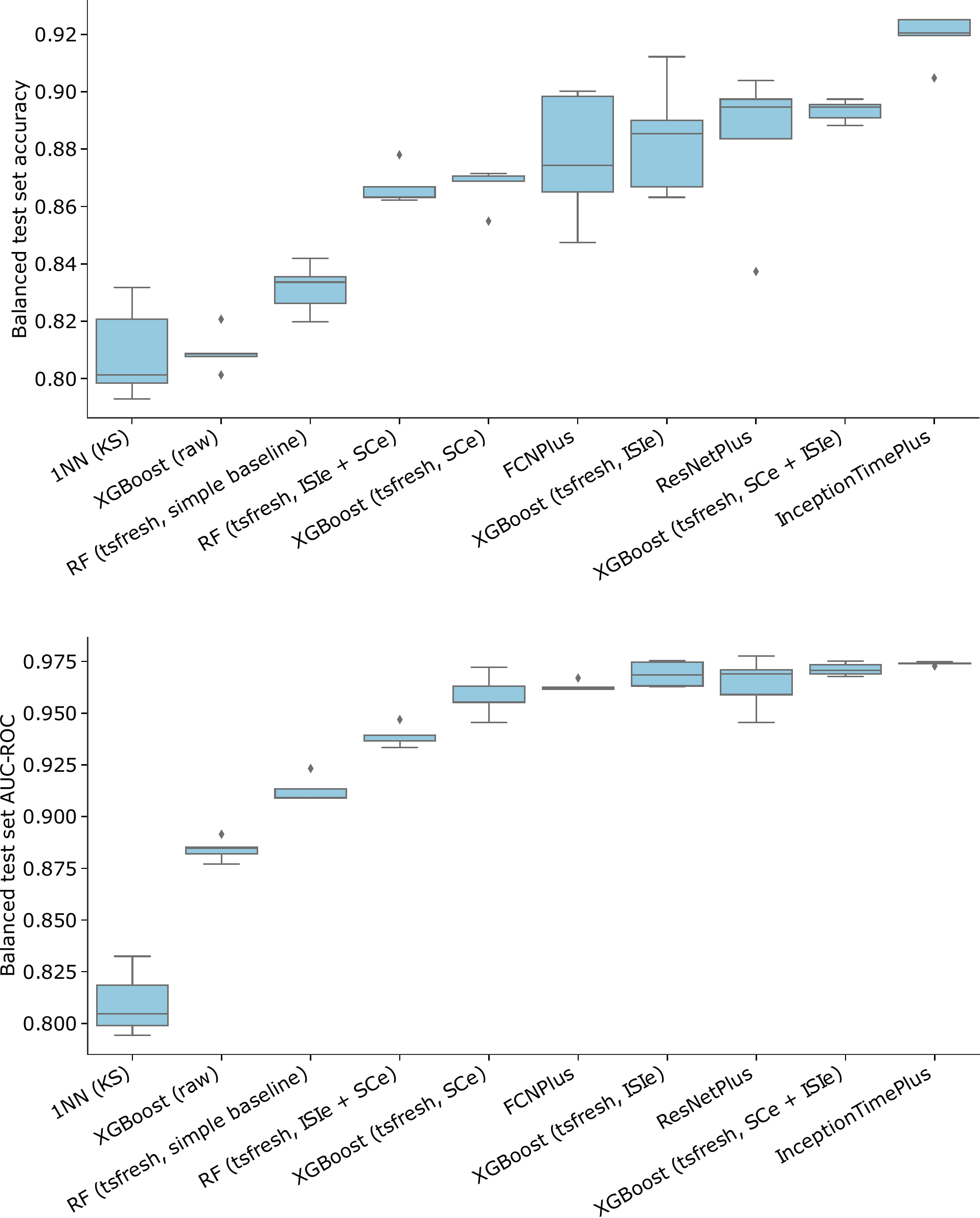}
    \caption{Spike train classification metric values (top panel - accuracy, bottom panel - AUC-ROC) for the retinal neuron activity dataset on a range of models. The task is defined as binary classification of the stimulus type (``white noise checkerboard'' or ``randomly moving bar''), with the test set balanced in class distribution by undersampling (that is, accuracy=0.5 corresponds to chance level). Models are ranked in ascending order of the median metric value. The ``simple baseline'' model tag corresponds to spike trains encoded with 6 basic distribution statistics, the ``raw'' tag implies that the model has been directly trained on ISI time-series data without feature extraction. The ``tsfresh'' tag corresponds to encoding with the full set of time-series features. ``ISIe'' stands for interspike-interval encoding of the spike train, ``SCe'' stands for spike-count encoding. ``ISIe + SPe'' means that feature vectors corresponding to both types of encoding are concatenated. InceptionTimePlus, FCNPlus, and ResNetPlus refer to implementations in the PyTorch-based \textit{tsai} package.}
    \label{fig:figure2}
\end{figure*}

\begin{table*}[!ht]
\centering
\begin{adjustwidth}{-2.25in}{0in}
{\color{black}\begin{tabularx}{\linewidth}{|l|X|X|X|}
\hline
{} & {Interspike interval encoding (ISIe)} & {Spike count encoding (SCe)} & {Interspike interval + spike count encoding (ISIe + SCe)} \\
\hline 
{Logistic Regression} & {{ \bf 84.10 $\pm$ 1.55}} & {75.05 $\pm$ 1.17} & {84.01 $\pm$ 0.35} \\ \hline 
{Random Forest} & {86.14 $\pm$ 0.77} & {85.86 $\pm$ 0.80} & {{ \bf 86.32 $\pm$ 0.65}} \\ \hline
{Extra Trees} & {85.58 $\pm$ 0.36} & {86.14 $\pm$ 1.02} & {{ \bf 86.23 $\pm$ 1.01}} \\ \hline 
{XGBoost} & {87.06 $\pm$ 1.98} & {88.54 $\pm$ 0.70} & {{ \bf 89.46 $\pm$ 0.37}} \\ \hline 

\end{tabularx}}
\vspace{1mm}
\caption{{\color{black}Balanced test set accuracy values on the retina dataset with different classifiers trained on tsfresh feature representations obtained from the (i) interspike interval encoding of the spike trains, (ii) the spike count encoding of the spike trains, (iii) combined interspike interval + spike count encoding.}}
\label{tab:table4}
\end{adjustwidth}
\end{table*}

\begin{table*}[!ht]
\centering
\begin{adjustwidth}{-2.25in}{0in}
{\color{black}\begin{tabularx}{\linewidth}{|l|X|X|}
\hline
{} & {Log-transform + z-score normalization} & {No preprocessing} \\
\hline 
{Logistic Regression on tsfresh features} & {0.6743} & {0.3012} \\ \hline 
{Random Forest on tsfresh features} & {0.7293} & {0.6009} \\ \hline
{XGBoost on tsfresh features} & {0.7526} & {0.6188} \\ \hline 
{InceptionTimePlus} & {0.8274} & {0.4145} \\ \hline 
{ResNetPlus} & {0.7723} & {0.3947} \\ \hline 
{FCNPlus} & {0.7111} & {0.4628}  \\ \hline 
{XceptionTimePlus} & {0.7959} & {0.0000 (not converged)}  \\ \hline 

\end{tabularx}}
\vspace{1mm}
\caption{Classifier performance degradation in case when no preprocessing for ISI-encoded data is used compared to our standard preprocessing pipeline. Cohen's kappa score for a range of models in the retinal neuron activity task is reported.}
\label{tab:table_prepro}
\end{adjustwidth}
\end{table*}

We evaluated the performance of nearest-neighbor models with different distance metrics on the retinal stimulus classification task, results are shown in Table \ref{tab:table1}. The results in presented in Table \ref{tab:table1} were obtained without performing undersampling to balance the class distribution, hence we looked at imbalance-robust metrics such as Cohen's kappa and geometric mean scores. We found that the nearest neighbor model with the DTW distance is amongst the best-performing ones, but is still outperformed by the 1-NN model with the Kolmogorov-Smirnov (KS) distance, suggesting that differences in ISI distributions contain significant discriminative information helpful for the classification task at hand. We further include the results obtained with the 1-NN KS-distance model as a baseline to compare against other methods.

\begin{table*}[!ht]
\centering
\begin{adjustwidth}{-2.25in}{0in}
{\color{black}\begin{tabularx}{\linewidth}{|l|X|X|}
\hline
{} & {Cohen's kappa score} & {Geometric mean score}  \\
\hline 
{InceptionTimePlus} & {{\bf 0.8274}} & {0.9051} \\ \hline 
{ResNetPlus} & {0.7723} & {0.8941} \\ \hline 
{FCNPlus} & {0.7111} & {0.8857}  \\ \hline 
{XceptionTimePlus} & {0.7959} & { {\bf 0.9271}}  \\ \hline 

{XGBoost (tsfresh, ISIe + SCe)} & {0.7799} & {0.9082}  \\ \hline 
{Random Forest (tsfresh, ISIe + SCe)} & {0.7018} & {0.8751} \\ \hline 

{XGBoost (tsfresh, ISIe)} & {0.7526} & {0.9005}  \\ \hline 
{Random Forest (tsfresh, ISIe)} & {0.7293} & {0.8853} \\ \hline 

{XGBoost (tsfresh, SCe)} & {0.6830} & {0.8306}  \\ \hline 
{Random Forest (tsfresh, SCe)} & {0.6100} & {0.7729}  \\ \hline 

{XGBoost (tsfresh, simple baseline)} & {0.5408} & {0.8287}  \\ \hline 
{Random Forest (tsfresh, simple baseline)} & {0.5160} & {0.8170}  \\ \hline 

{XGBoost (raw)} & {0.5426} & {0.7840}  \\ \hline 
{1NN (KS distance)} & {0.5464} & {0.8316}  \\ \hline 

\end{tabularx}}
\vspace{1mm}
\caption{{\color{black} Spike train classification metric values (for imbalance-robust metrics) for the retinal neuron activity dataset on a range of models. The ``simple baseline'' model tag corresponds to spike trains encoded with 6 basic distribution statistics, the ``raw'' tag implies that the model has been directly trained on ISI time-series data without feature extraction. The ``tsfresh'' tag corresponds to encoding with the full set of time-series features. ``ISIe'' stands for interspike-interval encoding of the spike train, ``SCe'' stands for spike-count encoding. ``ISIe + SPe'' means that feature vectors corresponding to both types of encoding are concatenated. InceptionTimePlus, FCNPlus, ResNetPlus and XceptionTimePlus and refer to implementations in the PyTorch-based \textit{tsai} package.}}
\label{tab:table2}
\end{adjustwidth}
\end{table*}

\begin{table*}[!ht]
\centering
\begin{adjustwidth}{-2.25in}{0in}
{\color{black}\begin{tabularx}{\linewidth}{|l|X|X|X|}
\hline
{} & {Retinal stimulus (white noise vs. moving bar)} & {fcx-1 WAKE/SLEEP state prediction} & {Allen Cell Types SST/VIP INs} \\
\hline 
{InceptionTimePlus} & {{\bf 92.05 $\pm$ 0.84}} & {76.82 $\pm$ 1.07} & {75.98 $\pm$ 1.76} \\ \hline 
{ResNetPlus} & {89.46 $\pm$ 2.68} & {77.01 $\pm$ 0.79} & {77.17 $\pm$ 2.61} \\ \hline 
{FCNPlus} & {87.43 $\pm$ 2.24} & {{ \bf 78.02 $\pm$ 0.58}} & {{ \bf 77.95 $\pm$ 2.85}} \\ \hline 
{XceptionTimePlus} & {86.41 $\pm$ 3.13} & {77.22 $\pm$ 0.83} & {{\bf 77.95 $\pm$ 1.79}} \\ \hline 

{XGBoost (tsfresh, ISIe + SCe)} & {89.46 $\pm$ 0.37} & {75.49 $\pm$ 0.21} & {68.50 $\pm$ 3.44} \\ \hline 
{Random Forest (tsfresh, ISIe + SCe)} & {86.32 $\pm$ 0.65} & {76.49 $\pm$ 0.29} & {66.92 $\pm$ 1.83} \\ \hline 

{XGBoost (tsfresh, ISIe)} & {88.54 $\pm$ 1.97} & {76.03 $\pm$ 0.73} & {69.68 $\pm$ 4.91} \\ \hline 
{Random Forest (tsfresh, ISIe)} & {86.14 $\pm$ 0.77} & {76.62 $\pm$ 0.46} & {66.14 $\pm$ 6.63} \\ \hline 

{XGBoost (tsfresh, SCe)} & {87.06 $\pm$ 0.67} & {73.79 $\pm$ 0.57} & {61.42 $\pm$ 2.34} \\ \hline 
{Random Forest (tsfresh, SCe)} & {85.86 $\pm$ 0.80} & {75.21 $\pm$ 0.46} & {64.96 $\pm$ 0.98} \\ \hline 

{Random Forest (tsfresh, simple baseline)} & {83.36 $\pm$ 0.86} & {74.00 $\pm$ 0.36} & {63.39 $\pm$ 3.45} \\ \hline 
{XGBoost (tsfresh, simple baseline)} & {82.99 $\pm$ 1.31} & {71.39 $\pm$ 0.60} & {59.45 $\pm$ 3.16} \\ \hline  

{XGBoost (raw)} & {80.87 $\pm$ 0.70} & {71.97 $\pm$ 0.32} & {63.78 $\pm$ 2.60} \\ \hline 
{1NN (KS distance)} & {80.13 $\pm$ 1.64} & {67.31 $\pm$ 1.44} & {61.02 $\pm$ 2.40} \\ \hline 

\end{tabularx}}
\vspace{1mm}
\caption{{\color{black} Spike train classification accuracy values for different datasets on a range of models. The reported accuracy is measured on balanced test sets to mitigate class imbalance, median value and standard deviation in percent are shown. Model names correspond to the same ones from Table 2.}}
\label{tab:table3}
\end{adjustwidth}
\end{table*}

\subsubsection*{Hand-crafted feature extraction + classification models}
\label{sec:tsfresh_inh_exc}

The kNN results clearly suggest that characteristics of the interspike-interval distribution of the given spike train are predictive of the category label in our classification task. At the same time, one would expect the temporal (sequential) information contained in the spike train also has certain predictive power. A straightforward way to incorporate both types of features in the model is to build a corresponding vector embedding of the spike train time series. An efficient way to do so is to use a set of hand-crafted time-series features, like for example the set of 779 features provided in the \textit{tsfresh} Python package. In order to compute vector embeddings for the spike trains in the training and testing datasets, one has to convert spike times into a time series, which in principle could be done using either an interspike-interval encoding (the time-series is the sequence of ISIs) or a spike-count encoding (time is binned and spike counts in each time bin comprise the time series). The latter type of encoding depends on an additional hyperparameter which is the size of the time bin while ISI encoding is parameter-free. We tested both types of spike-train encoding for our task and observed that models trained using the ISI-encoding of spikes generally perform better than the ones using binned spike counts. Furthermore, we found that combining features corresponding to both encoding types leads to better performance compared to using a single encoding scheme (see Table \ref{tab:table4} and Fig \ref{fig:figure2}). Crucially, we found that not applying the logarithmic transform to the ISI-encoded data (which is heavy-tail distributed) leads to significant accuracy degradation for \textit{tsfresh}-based models as well as other model types such as deep neural networks (see Table \ref{tab:table_prepro} and Fig \ref{fig:figureS2}).

\begin{figure*} [!htbp]
    \centering
    \includegraphics[width=0.9\textwidth]{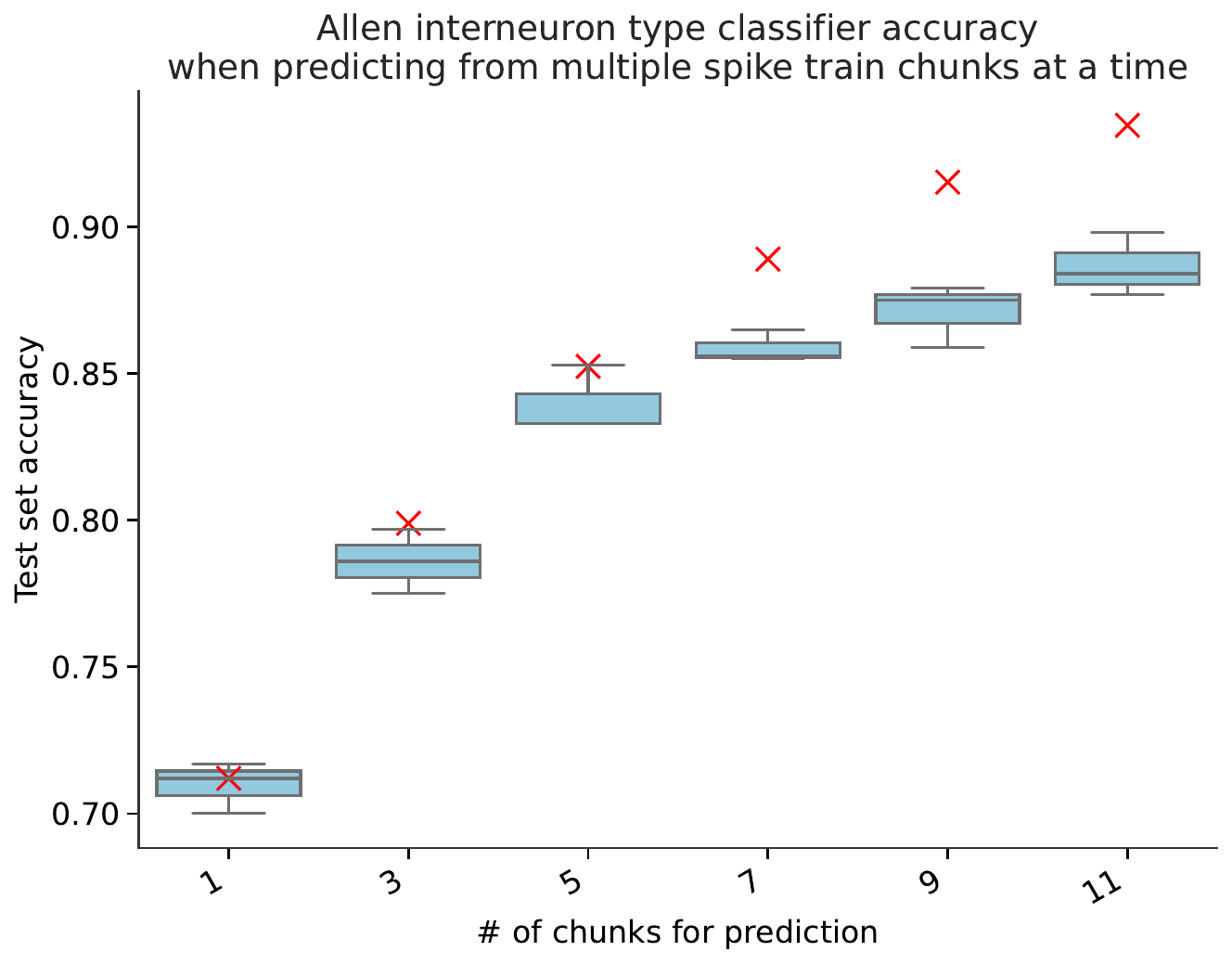}
    \caption{Classification accuracy for the Allen cell types VIP/SST interneuron classification task in the case of multiple randomly sampled same-class spike train chunks per prediction (with prediction done via majority voting). The model trained in these trials is a random forest classifier on the full set of \textit{tsfresh} features. The boxplots reflect the median accuracy and the variance between different train/test splits as done in the main text for the fcx-1 data set. The red crosses correspond to the theoretical estimate under the assumption of independently sampled spike train chunks.}
    \label{fig:figure3}
\end{figure*}

For each spike-train encoding type, we computed the 779-dimensional \textit{tsfresh} time-series embeddings independently for each sample in the training and testing datasets (no statistic aggregation across samples is performed). We then performed simple pre-processing steps by (i) removing low-variance features from the embedding (features $f$ satisfying std$(f)/($mean$(f) + \varepsilon) < \theta$ with $\theta=0.2$ and $\varepsilon=10^{-9}$ were removed) and (ii) performing z-score normalization for each feature using mean and variance statistics collected over the training dataset. 
Note that since the number of spikes in each data sample is fixed and interspike intervals are highly variable, the number of time bins also becomes variable from sample to sample. The fixed-size \textit{tsfresh} embeddings can nevertheless be computed since they are applicable to variable-length time series.

We then trained classification models on the resulting spike-train vector embeddings. We chose a representative set of classification models comprising (i) a linear model, namely logistic regression with an $l_2$ regularization penalty, and (ii) several types of tree-based ensembles: a random forest classifier (via the scikit-learn's RandomForestClassifier implementation), randomized decision trees (via the scikit-learn's ExtraTreesClassifier), and a gradient boosted decision tree (GBDT) ensemble (via the XGBoost implementation). The classifier hyperparameter values we used are specified in S1 Text. We have not included validation metric values of the logistic regression models in figures and tables because we found that linear models always perform worse compared to decision tree ensembles.

Classification results obtained with the described \textit{tsfresh}-based approach are presented in Fig \ref{fig:figure2} and Tables \ref{tab:table2}, \ref{tab:table3}. Note that in cases where we report balanced test set metrics (accuracy and AUC-ROC) we have performed class balancing via undersampling on the dataset and also have randomly sampled 70\% of the training set data over several trials to estimate the variance in validation metrics. In cases when we don't perform any undersampling, we report the values of Cohen's kappa and geometric mean score metrics. 

We were able to reach significant performance levels ($>0.88$ accuracy, $>0.95$ AUC-ROC) with our best \textit{tsfresh}-based models on the binary retinal stimulus classification task (``randomly moving bar'' vs. ``white noise checkerboard''). To make better sense of these metric values, we compared our \textit{tsfresh}-based models against two simple baselines: (a) a logistic regression model on ISI-encoded spike-trains represented by 6 basic statistical features -- the mean, median, minimum and maximum ISI values, the standard deviation and the absolute energy of the ISI-sequence (the mean of squared ISI values) and (ii) an XGBoost model trained directly on ``raw'' ISI-encoded spike trains.

The best-performing model using the 6 basic features of the ISI time series got a median balanced test set accuracy of 83.36, while training an XGBoost model directly on the ISI time series gave an accuracy of 80.86. Using full \textit{tsfresh} embeddings on ISI time-series boost the accuracy to 88.53 with the best model (XGBoost). We generally found that using the binned spike count encoding of spike trains works worse with \textit{tsfresh} compared to the ISI encoding (see Table \ref{tab:table4}), while combining feature vectors obtained from both encodings results in improved accuracy (89.46 for the XGBoost model on the retina dataset). 

We also evaluated the performance of state-of-the-art deep learning models in our retinal stimulus classification tasks, using implementations from the \textit{tsai} package \cite{tsai}. Accuracy results for a range of convolutional architectures (FCN, InceptionTime, XceptionTime, ResNet) are shown in Tables \ref{tab:table2}, \ref{tab:table3}. We found that convolutional neural nets generally outperform manual-feature-based methods, although by a relatively small margin both in cases of class-balanced and imbalanced datasets.  

We then performed the same pre-processing steps for the WAKE/SLEEP and VIP/SST datasets as for the retinal stimulus classification dataset. The rolling window of size equal to 200 ISIs and a stride of 100 ISIs for the WAKE/SLEEP data produced a dataset of 24363 training samples (from 78 neurons) and 10634 testing samples (from 35 neurons) with average target values of 0.3796 and 0.3211 in the training and testing sets, correspondingly. A rolling window of size equal to 50 ISIs and a stride of 20 ISIs for VIP/SST interneuron data produced a dataset of 2690 training samples and 1217 testing samples, with mean target values of 0.6026 and 0.8504, correspondingly. 

We observed similar trends both for the WAKE/SLEEP state and VIP/SST interneuron classification tasks, shown in Table \ref{tab:table3}. The best performing non-deep-learning models were found to be \textit{tsfresh}-based ones using the combined ISI and spike count encodings of the underlying spike trains. Convolutional neural networks were found to outperform the classical models, sometimes by a considerable margin (i.e. the Allen cell types dataset).

The base task we consider for all benchmark datasets is classification given an individual spike train chunk (part of a single-neuron recording). However, prediction performance can be improved by aggregating predictions from spike trains of several neurons or from several chunks of a large single-neuron spike train. If simultaneous spike train recordings from several neurons are available, one could formulate the learning problem as multivariate time-series classification; furthermore, most of the learning algorithms considered in this study can be extended to handle multivariate time-series as input data (e.g. using several independent input channels in CNN models, concatenating \textit{tsfresh} embedding vectors from multiple samples, etc.). In our univariate time-series (single-neuron spike train) formulation, the simplest way to extend the results to a multi-spike train setting is to ensemble the predictions of a single model made on multiple samples corresponding to the same class via majority voting. If we assume that the spike-trains chunks are randomly sampled from the whole test set in this process, the optimistic estimate for accuracy improvement with the number of spike train chunks $N_{\text{chunks}}$ would be
\begin{equation} \label{major_vote_accuracy}
    \mu = 1 - \sum_{i=0}^{\lfloor N_{\text{chunks}} / 2 \rfloor} C_{N_{\text{chunks}}}^i p^i (1-p)^{N_{\text{chunks}}-i}
\end{equation}
where $\mu$ is the probability that the majority vote prediction is correct, $p$ is the probability of a single classifier prediction being correct (single spike train prediction accuracy), $N_{chunks}$ is the number of predictions made. We found that the empirical values of accuracy improvement are close to the optimistic analytical estimate (\ref{major_vote_accuracy}) in both cases when the spike train chunks are sampled from different neurons and from a large spike train of a single neuron (see Fig \ref{fig:figure3}).

Being able to estimate feature importance ranks from trained decision tree ensembles allows us to detect the most discriminating features of ISI time series. We have analyzed the feature importance scores for \textit{tsfresh} features on all three datasets and found that the set of discriminative features is significantly overlapping for the three tasks (in terms of feature types) and includes both features that depend on the characteristics of the ISI distribution as well as features that characterize the temporal structure of ISI time-series (see S1 Text).

\begin{figure}
    \centering
    \includegraphics[width=\textwidth]{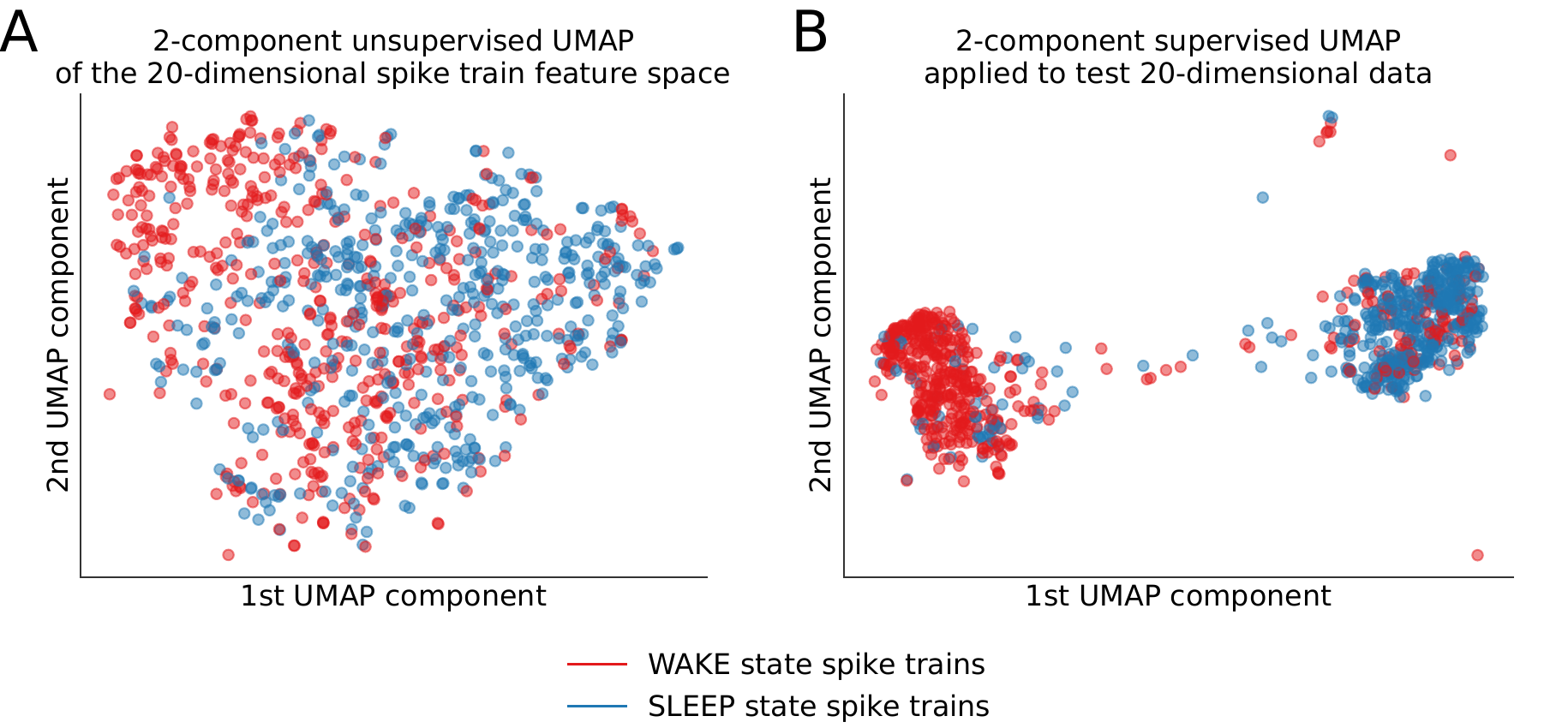}
    \caption{Spike train feature embeddings for WAKE (points marked red) vs. SLEEP (points marked blue) activity states of the neural circuit. Two-dimensional embeddings of the (20-dimensional) selected-\textit{tsfresh}-feature space using (A) unsupervised UMAP and (B) supervised UMAP embedding algorithms for spike trains corresponding to WAKE vs. SLEEP activity states. 
    }
    \label{fig:figure4}
\end{figure}

To visualize class separation for the WAKE vs. SLEEP state spike trains from the fcx-1 dataset as point clouds in two dimensions, we took the top-20 importance \textit{tsfresh} features identified during the feature selection procedure. We then used dimensionality reduction techniques on this reduced 20-dimensional dataset to visualize the structure of the data with respect to the WAKE/SLEEP state labels of the series. Results are shown in Fig \ref{fig:figure4} for two Uniform Manifold Approximation and Projection, UMAP \cite{mcinnes2018umap-software} low-dimension embedding algorithms. In all cases, classes cannot be linearly separated in two-dimensional embedding spaces, however, there is a separation of a large fraction of the points of the WAKE and SLEEP state classes.

We conclude that both the convolutional neural networks and the hand-crafted feature engineering approach combined with strong tree-based learning models set a strong baseline for spike train classification for all of the three studied tasks.

\subsubsection*{Unsupervised spike train temporal structure recognition}

The spike train temporal structure recognition task is defined as follows: for a set of spike train activity data, we generate a binary classification task by producing an additional category of spiking data consisting of spike trains from the original dataset with a certain transformation applied to them. We consider the following spike train transformations: (i) ISI shuffling inside the spike train (random shuffling applied to the ISI time series), (ii) reversing the ISI time series and (iii) adding spike timing jitter sampled from the truncated normal distribution to the time series. Note that the first two transformation types do not change the value distribution of the time series, only its temporal structure (the exact ordering of the interspike intervals of the spike train). Hence, if it is possible to construct a classification model capable of distinguishing between the two activity classes (original spiking activity versus the transformed one), then one could say that the model has learned to detect the temporal structure in the time series. In the case of classifiers trained on \textit{tsfresh} feature vectors, the classification accuracy metrics obtained can be thought of as measures of the amount of temporal structure contained in the spike trains (to the extent encoded in \textit{tsfresh} features). The classification results (AUC-ROC values) for different base spiking data and different transforms are shown in Table \ref{tab:table_unsuper}. 
Notably, one could observe higher AUC-ROC values for the retinal ganglion cell spiking data in the case of the randomly moving bar stimulus as compared to the white noise checkerboard stimulus for all of the three transforms considered. The same difference in accuracy values is observed for the fcx-1 dataset whereby classification AUC-ROC value is higher for all of the three transforms when the SLEEP state is used as the base spiking dataset as opposed to the WAKE state. The accuracy values that we observe for the temporal structure recognition tasks are above the chance level in most cases, with low values for fcx-1 WAKE-state data in case of reverse and noise transforms.

\begin{table*}[!ht]
\centering
\begin{adjustwidth}{-2.25in}{0in}
\begin{tabularx}{\linewidth}{|X|X|X|X|}
\hline
{} & {Reverse transform} & {Shuffling transform} & {Noise transform} \\
\hline 
{Retinal ganglion cells (randomly moving bar input)} & {0.8178} & {0.9052} & {0.70266} \\ \hline 
{Retinal ganglion cells (white noise checkerboard input)} & {0.8693} & {0.9273} & {0.9360} \\ \hline 
{fcx-1 WAKE spike trains} & {0.5881} & {0.9366} & {0.5418} \\ \hline 
{fcx-1 SLEEP spike trains} & {0.7611} & {0.9522} & {0.5963} \\ \hline 
\end{tabularx}
\vspace{1mm}
\caption{Test set AUC-ROC values for the unsupervised temporal structure recognition tasks for different base spiking datasets and different transforms. A random forest classifier model (see S1 Text for the hyperparameter values used) was used in all of the above experiments.}
\label{tab:table_unsuper}
\end{adjustwidth}
\end{table*}

\section*{Discussion}

In this work, we have introduced a diverse neuronal spike train classification benchmark to evaluate neural decoding algorithms. The benchmark consists of several single-neuron spike train prediction tasks spanning stimulus type prediction, neuron type identification, and animal behavioral state prediction. The interneuron type classification task (SST vs. VIP interneurons) is shown to be non-trivial due to the firing patterns of these two neuron types being quite similar in properties (as opposed to, for instance, PV interneurons, which typically have significantly higher firing rates compared to SST/VIP neurons as well as pyramidal cells). The other two tasks relate the spiking activity of individual neurons to the global state of the underlying neural circuit, which is in one of the cases stationary during a considerable time period (fcx-1 dataset) and the other one could be viewed as a transient stimulus-driven one (retinal dataset). In both cases, we have demonstrated that individual neuronal spike trains contain information related to the global state of the neural circuit and this information can be decoded (from a relatively small spike train) if appropriate time-series learning models are used. Extensive experiments on several datasets that we have conducted imply that not only ISI value distribution is important for global state identification but also the temporal information contained in the spike trains, that is, features related to the exact sequences of interspike intervals in neural firing. We have identified groups of features highly informative for neural decoding tasks and established that this feature encoding combined with strong supervised learning algorithms such as gradient-boosted tree ensembles establishes a strong baseline on the proposed benchmark that performs on par with state-of-the-art deep learning approaches. While deep learning models are generally found to outperform the feature-extraction-based approaches by a certain margin, the feature-based approach could be favored for its easier interpretability. Extracting the spike train features important for classification would require extra steps in case deep learning models are used and can be done in a more straightforward way using the feature-based approach (e.g. to determine the frequency bands most important for the classification task). Another issue for the usage of deep learning models in the neural decoding context could be the limited amount of data available. While for the tasks proposed in spikebench with thousands of samples in each of the datasets deep learning is found to perform well, the feature-based approaches might be more robust when training on just hundreds of samples.

The feature-based approaches were also found to be robust to relatively low time-series sample duration. We have shown that significantly large accuracy values can be obtained on all of the proposed tasks using the hand-crafted feature encoding approach on single-neuron spike train chunks containing as low as 50 interspike intervals. We suggest that accuracy values can further be improved by hyperparameter search, model ensembling and test-time data augmentation. We propose that neural decoding models be evaluated on diverse and challenging tasks including the proposed benchmark (as well as regression tasks used to evaluate decoding models previously \cite{glaser2017machine,livezey2021deep}) in order to establish a sensible model performance ranking similar to what is done for computer vision and natural language understanding problems. We believe that this would drive further development of highly accurate neural decoding/neural activity mining approaches enabling their application in precision-critical tasks such as identifying pathological disease-related firing activity patterns in the brain.

\section*{Conclusion}

To summarize our contributions, we have proposed a challenging and diverse benchmark for \textit{individual cell} spike train classification to evaluate neural decoding models. 

We have shown that a classical machine learning baseline comprised of massive time-series feature extraction from different spike train encodings coupled with well-performing classification approaches such as gradient boosting produces results on par with deep learning models, although with deep neural nets still slightly outperforming the classical methods. Furthermore, we have shown that the firing of individual neurons contains information about the global state of the organism as well as information about the neuron type that can be decoded with machine learning approaches. This approach was further generalized to the unsupervised (self-supervised) setting, which helped reveal interesting structural properties of the spiking data we considered, in particular, the WAKE-state time-reversal invariance and spiking jitter robustness of the cortical activity in the fcx-1 dataset. The massive time-series feature engineering approach helped detect groups of time-series features that have discriminative power over a set of different tasks in our benchmark and might thus be useful in general neural decoding tasks.   


\clearpage

\section*{Supplementary materials.}

\section*{A. Classifier hyperparameter values.}

Listed below are hyperparameter values and implementation references for all of the classifier types we used. For more reference, see \href{https://github.com/lzrvch/spikebench/blob/master/paper_figures/tsfresh_zoo.py}{the example script}.

\begin{itemize}
\item Random Forest: \href{https://scikit-learn.org/stable/}{sklearn} implementation, $n\_estimators=500$, $max\_depth=10$.
\item Extra Trees Classifier: \href{https://scikit-learn.org/stable/}{sklearn} implementation, $n\_estimators=500$, $max\_depth=None$ (no limit on depth).
\item Logistic Regression: \href{https://scikit-learn.org/stable/}{sklearn} implementation, $l_2$ penalty, $C = 0.001$
\item XGBoost: \href{https://xgboost.readthedocs.io/en/latest/}{xgboost} implementation
\begin{itemize}
\item $max\_depth=8$
\item $learning\_rate=0.1$
\item $n\_estimators=500$
\item $objective=binary:logistic$
\item $booster=gbtree$
\item $gamma=0$
\item $min\_child\_weight=1$
\item $max\_delta\_step=0$
\item $subsample=0.7$
\item $colsample\_bytree=1$
\item $colsample\_bylevel=1$
\item $colsample\_bynode=1$
\item $reg\_alpha=0$
\item $reg\_lambda=1$
\item $scale\_pos\_weight=1$
\item $base\_score=0.5$
\end{itemize}
\item FCN, InceptionTime, XceptionTime, ResNet: \href{https://timeseriesai.github.io/tsai/}{tsai} implementation

\begin{itemize}
\item $epochs=200$
\item $max\_lr=0.1$
\item $optimizer=sgd$
\item $weight\_decay=1e-4$
\item $batch\_size=128$
\item $lr\_schedule=cosine$
\item $best\_model=$ with the largest $cohen\_kappa$ or $accuracy$ on the validation set
\end{itemize}
\end{itemize}

We have looked at how the performance of a CNN is robust to the above training hyperparameters using the XceptionTime architecture on the retinal stimulus classification dataset (see Table \ref{tab:tables1} in S1 Text).

\renewcommand{\thetable}{A}
\setcounter{table}{0}

\begin{table*}[!ht]
\centering
\begin{tabularx}{\linewidth}{|l|X|}
\hline
{Varied hyperparameters w.r.t. default setting} & {Geometric mean score} \\
\hline 
{Default setting} & {0.9270} \\ \hline 
{Peak lr = 0.01} & {0.8978} \\ \hline 
{Adam optimizer, peak lr = 0.01} & {0.9048} \\ \hline 
{One cycle lr schedule} & {0.9095} \\ \hline 
{100 training epochs} & {0.9204} \\ \hline 
{Batch size = 256} & {0.9198} \\ \hline 
\end{tabularx}
\vspace{1mm}
\caption{Geometric mean score obtained for the XceptionTime architecture trained on the retinal stimulus classification dataset with alterations in training hyperparameters.}
\label{tab:tables1}
\end{table*}

We found that generally the CNN performance is not significantly affected by the changes in the main hyperparameters, being relatively robust with respect to these changes. We do, however, expect that a thorough hyperparameter search (including search over the architecture's parameters) could result in a significant improvement in performance.

\clearpage
\section*{B. Discriminative tsfresh features.}

In order to select the important groups of discriminative features, we trained several random forest classification models (with different random seeds) on each dataset and used the feature importance scores extracted from the classifiers to rank the \textit{tsfresh} features. We have used the mean scores averaged over different datasets to identify the features that are discriminative for spiking data across different scenarios. According to this feature ranking procedure, the following groups of \textit{tsfresh} features are selected (see also Fig \ref{fig:figureS2} in S1 Text):

\begin{itemize}
\item $median$, $kurtosis$, $quantile\_q$ -- simple statistics of the ISI value distribution in the series like the median ISI value, $q$ quantiles and kurtosis of the ISI value distribution
\item $change\_quantiles$ -- this feature is calculated by fixing a corridor of the time series values (defined by lower and higher quantile bounds, $q_l$ and $q_h$, which are hyperparameters), then calculating a set of consecutive change values in the series (differencing) and then applying an aggregation function (mean or variance). Another boolean hyperparameter $is\_abs$ determines whether absolute change values should be taken or not.
\item $fft\_coefficient$ -- absolute values of the fast Fourier transform coefficients (individual coefficient values and aggregates).
\item $entropy$ -- values of the sample entropy, the approximate entropy and the binned entropy of the power spectral density of the time series.

\item $agg\_linear\_trend$ -- features from linear least-squares regression (standard error in particular) for the values of the time series that were aggregated over chunks of a certain size (with different aggregation functions like min, max, mean and variance). Chunk sizes vary from 5 to 50 points in the series.
\end{itemize}

\clearpage

\renewcommand{\thefigure}{A}
\setcounter{figure}{0}

\begin{figure*} [!htbp]
    \centering
    \includegraphics[width=1.\textwidth]{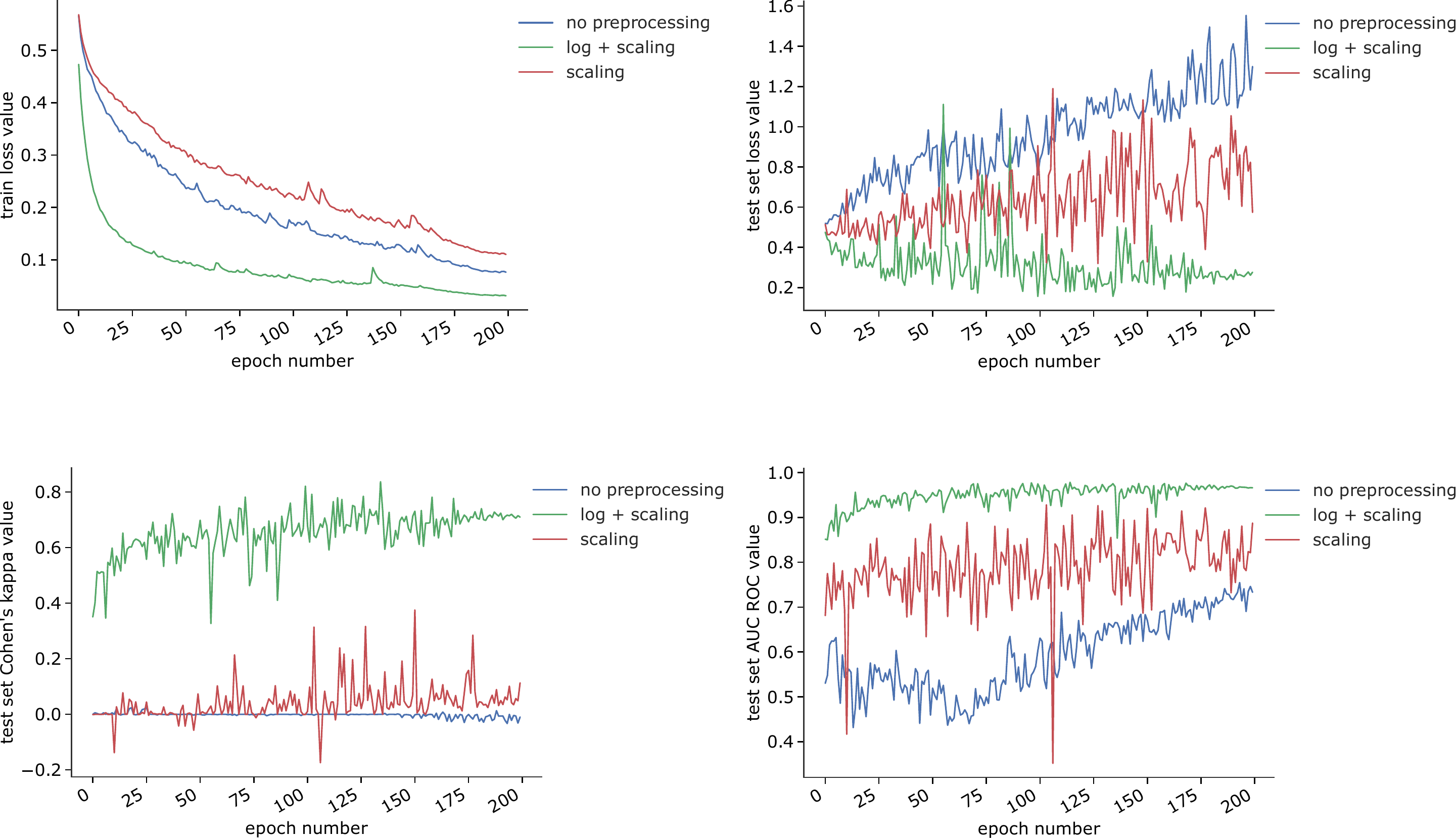}
    \caption{\textcolor{black}{Metric value evolution during training of an XceptionTime model on the retina dataset with different data preprocessing strategies: blue -- no preprocessing, original ISI sequences are used as input; red -- standard scaling is performed before feeding the time series to the CNN; green -- log-transform ($f(x)=\log(x+1)$) and standard scaling is applied to the input time series. Top left -- training set loss evolution, top right -- testing set loss evolution, bottom left -- Cohen's kappa score evolution on the test set, bottom right - test set AUC-ROC evolution during training.
    One can observe diverging test set loss in cases of no preprocessing or just standard scaling, at the same time training metrics are well-behaved when the log transform is applied to the data.}}
    \label{fig:figureS3}
\end{figure*}

\clearpage

\renewcommand{\thefigure}{B}
\setcounter{figure}{0}

\begin{figure*} [!htbp]
    \centering
    \includegraphics[width=1.\textwidth]{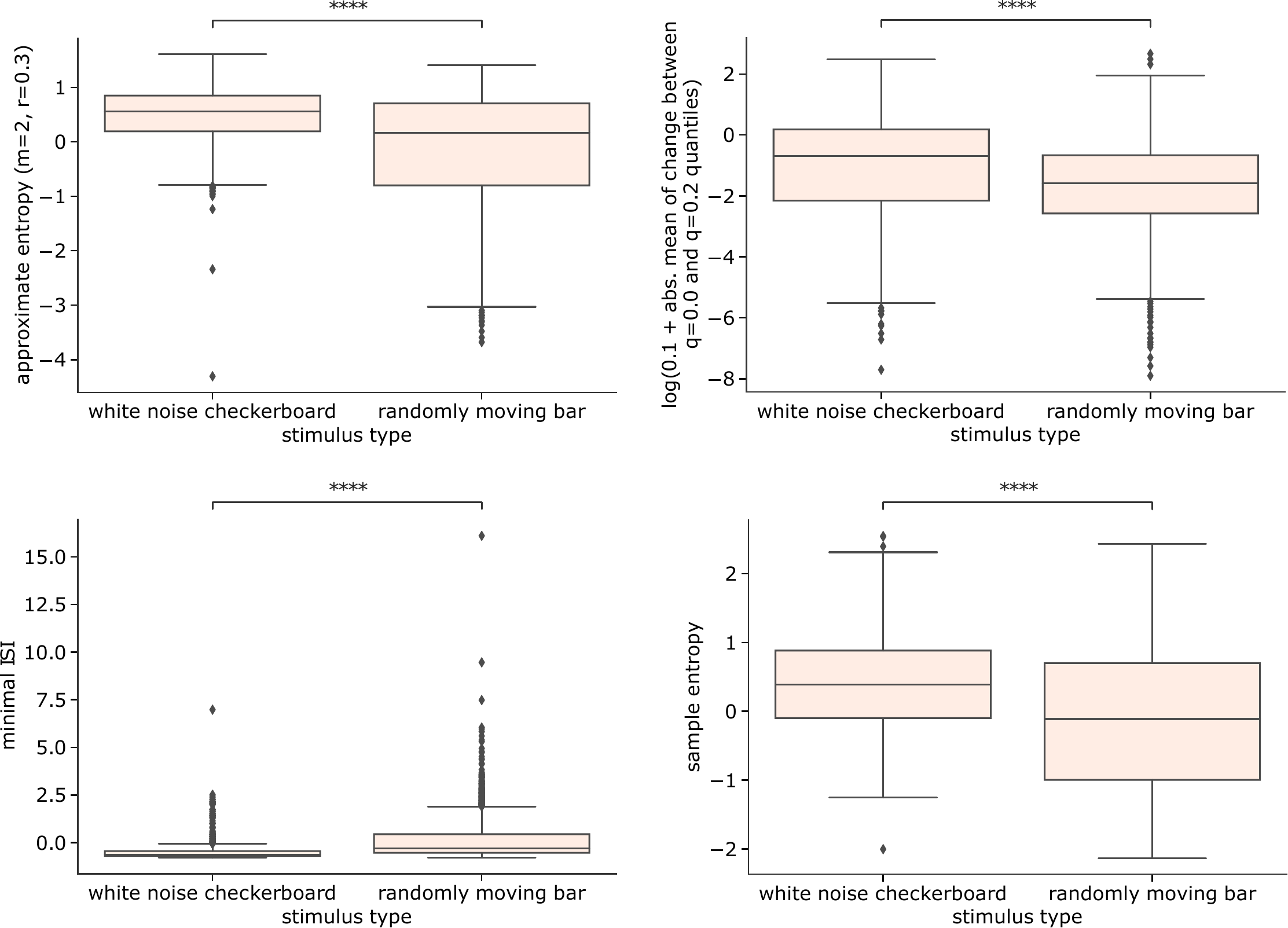}
    \caption{Boxplots of \textit{tsfresh}-extracted feature distributions for features with high discriminative power as detected by the trained decision tree ensemble classifiers in the retinal stimulus type prediction task. Two-sided Mann-Whitney-Wilcoxon test with Bonferroni correction is performed to assess statistical significance; **** denotes $p < 1e-4$.}
    \label{fig:figureS2}
\end{figure*}

\end{document}